\documentclass{article}

\usepackage{PRIMEarxiv}
\usepackage[utf8]{inputenc} % allow utf-8 input
\usepackage[T1]{fontenc} % use 8-bit T1 fonts
\usepackage{setspace} 
\usepackage{hyperref}       % hyperlinks
\usepackage{url}            % simple URL typesetting
\usepackage{booktabs}       % professional-quality tables
\usepackage{amsfonts}       % blackboard math symbols
\usepackage{nicefrac}       % compact symbols for 1/2, etc.
\usepackage{microtype}      % microtypography
\usepackage{lipsum}
\usepackage{fancyhdr}       % header
\usepackage{graphicx}       % graphics
\graphicspath{{media/}}     % organize your images and other figures under media/ folder
\usepackage{tablefootnote}
\usepackage{array}
\usepackage[longtable]{multirow}
\usepackage{longtable}
\usepackage{vcell}
\usepackage{enumitem}
\usepackage{authblk}  
\usepackage{booktabs}

\title{ Real-time Geoinformation Systems to Improve the Quality, Scalability, and Cost of Internet of Things for Agri-environment Research
}

\author[1,2-*]{Bryan C. Runck \thanks{Corresponding author - Bryan Runck, runck014@umn.edu}}
\author[1-]{Bobby Schulz}
\author[1]{Jeff Bishop}
\author[3]{Nathan Carlson}
\author[4]{Bryan Chantigian}
\author[6]{Gary Deters}
\author[3]{Jesse Erdmann}
\author[12]{Patrick M. Ewing}
\author[1,2]{Michael Felzan}
\author[]{Xiao Fu}
\author[24]{Jan Greyling}
\author[20]{Christopher J. Hogan}
\author[6]{Andrew Hollman}
\author[1,5]{Ali Joglekar}
\author[1]{Kris Junker}
\author[21]{Michael Kantar}
\author[]{Lumbani Kaunda}
\author[13,15]{Mohana Krishna}
\author[3]{Benjamin Lynch}
\author[8,10]{Peter Marchetto}
\author[]{Megan Marsolek}
\author[16]{Troy McKay}
\author[11]{Brad Morris}
\author[17]{Ali Rashid Niaghi}
\author[26]{Keerthi Pamulaparthy}
\author[1,5]{Philip Pardey}
\author[1]{Ann Piotrowski}
\author[25]{Christina Poudyal}
\author[3]{Tom Prather}
\author[7]{Barath Raghavan}
\author[23]{Maggie Reiter}
\author[1,2]{Lucas Rosen}
\author[]{Benjamin Salazar}
\author[22]{Andrew Scobbie}
\author[14]{Vasudha Sharma}
\author[1,3]{Kevin A. T. Silverstein}
\author[14]{Gurparteet Singh}
\author[14,18]{Jeff Strock}
\author[14]{Samikshya Subedi}
\author[19]{Evan Tang}
\author[1]{Gianna Turturillo}
\author[6]{Eric Watkins}
\author[22]{Blake Webster}
\author[27]{James Wilgenbusch}

\affil[-]{co-first authors}
\affil[1]{GEMS Informatics Center, University of Minnesota – Twin Cities}
\affil[2]{Department of Geography, Environment and Society, University of Minnesota – Twin Cities}
\affil[3]{Minnesota Supercomputing Institute, University of Minnesota }
\affil[4]{Department of Electrical Engineering, University of Minnesota – Twin Cities}
\affil[5]{Department of Applied Economics, University of Minnesota – Twin Cities}
\affil[6]{Department of Horticultural Science, University of Minnesota – Twin Cities}
\affil[7]{Department of Computer Science, University of Southern California}
\affil[8]{Sensing Inc}
\affil[10]{Conservify}
\affil[11]{Graphspan LLC}
\affil[12]{USDA-ARS Food Systems Research Unit, Burlington, VT 05405}
\affil[13]{Amazon}
\affil[14]{Department of Soil, Water and Climate, University of Minnesota – Twin Cities}
\affil[15]{Department of Computer Science and Engineering, University of Minnesota – Twin Cities}
\affil[16]{Departmnet of Agricultural Education, Communication and Marketing, University of Minnesota – Twin Cities}
\affil[17]{Benson Hill}
\affil[18]{Southwest Research and Outreach Center, University of Minnesota – Twin Cities}
\affil[19]{Code Weavers}
\affil[20]{Department of Mechanical Engineering, University of Minnesota – Twin Cities}
\affil[21]{University of Hawaii at Manoa, Honolulu, HI }
\affil[22]{Minnesota Agricultural Experiment Station, University of Minnesota – Twin Cities}
\affil[23]{Sunday}
\affil[24]{Stellenbosch AgroInformatics Initiative, Stellenbosch University}
\affil[25]{Ameriprise Financial Services, LLC}
\affil[26]{Optum}
\affil[27]{Research Computing, University of Minnesota - Twin Cities}

\makeatletter
\let\oldaffillist\AB@affillist
\renewcommand{\AB@affillist}{\begin{flushleft}\begin{spacing}{}\oldaffillist\end{spacing}\end{flushleft}}
\makeatother
 
\begin{document}
\maketitle
\nocite{*}
\vspace{80pt}
% \newpage

\begin{abstract}
With the increasing emphasis on machine learning and artificial intelligence to drive knowledge discovery in the agricultural sciences, spatial internet of things (IoT) technologies have become increasingly important for collecting real-time, high resolution data for these models. However, managing large fleets of devices while maintaining high data quality remains an ongoing challenge as scientists iterate from prototype to mature end-to-end applications. Here, we provide a set of case studies using the framework of technology readiness levels for an open source spatial IoT system. The spatial IoT systems underwent 3 major and 14 minor system versions, had over 2,727 devices manufactured both in academic and commercial contexts, and are either in active or planned deployment across four continents. Our results show the evolution of a generalizable, open source spatial IoT system designed for agricultural scientists, and provide a model for academic researchers to overcome the challenges that exist in going from one-off prototypes to thousands of internet-connected devices.
\end{abstract}

\keywords{ open source, geospatial, data interoperability, quality, internet of things, data quality, manufacturing, real-time, loggers}

\section{Introduction}
Artificial intelligence and machine learning are rapidly becoming core to the agricultural sciences \cite{basso_digital_2020}. These new approaches for innovating within food and agriculture require many advancements, from a shift in skills for researchers toward digital technologies to the development of reproducible and modular software targeted at agricultural applications. Fundamental to this shift toward digital agriculture is the need for large, labeled, clean, representative and analysis-ready datasets \footnote[1]{Here, we are describing quality controlled “big data” and rely on the 3V’s attributes of “large” or “big” - variety, velocity, and volume \cite{sagiroglu_big_2013}}. A growing ecosystem of workflows, tools, and platforms continue to develop in support of data-intensive agricultural research and innovation \cite{runck_digital_2022}. The intention of these technology ecosystems is to make data and workflows findable, accessible, interoperable, and reproducible \cite{wilkinson_fair_2016}, but work remains at the fundamental level of big and interoperable dataset generation.

Spatial internet of things (IoT) technologies are one of the many areas of active development across the digital agriculture sciences \cite{khanna_evolution_2019}. Spatial IoT in part involves the development of internet-connected sensors that can be ubiquitously embedded throughout agricultural systems where they collect geospatially and temporally referenced data. Through real-time data pipelines from sensors, it is possible to automate on-the-fly data quality and assurance to check for error-prone steps in data collection processes. Such systems have existed for some time in targeted application areas such as agrometeorological and hydroclimatic monitoring \cite{sawant_interoperable_2017}; however, general system design principles for multiple domains remain elusive. The promise of broad, cross-domain adoption of quality controlled, real-time spatial IoT is that scientists will be able to practically achieve a denser spatio-temporal resolution in sensing, which is expected to lead to better characterization, understanding, and prediction of agricultural systems that are intrinsically variable in space and time \cite{tao_review_2021}.

Such automated IoT pipelines are more common in individual domains of science and engineering, such as smart buildings \cite{minoli_iot_2017}, meteorology \cite{brock_oklahoma_1995}, and earth science \cite{hirschfeld_instrumentation_1985,hart_environmental_2006,hart_toward_2015}. Though progress has been made to realize the benefits of ubiquitous, real-time data streaming in agriculture \cite{tao_review_2021}, challenges remain in developing generalizable systems that are broadly transparent to scientific users, reliably produce high quality data, and are scalable both in terms of time and money costs \cite{mao_low-cost_2019}. 

A lack of practical systems maintenance, operation, and data quality assurance that generalize across domains hinders the broader adoption of spatial IoT technologies across the multiple disciplines encapsulated within the agricultural sciences. There are many reasons for this, including:  1) IoT in agriculture requires a larger geographic extent and greater spatial density than many other application areas, such as smart buildings or meteorology, and 2) agricultural environments are difficult for IoT due to both inherent difficulties (e.g. soil is difficult to transmit through and highly heterogeneous; large mobile machinery interacting with and around devices) and added logistical challenges (field operations for device management).

Here we focus on a subset of questions within the large domain of IoT in agriculture. How can we design spatial IoT systems so they can be (1) used for rapid prototyping of new hardware and software systems, (2) readily upscaled by agricultural scientists to large spatial extents and high densities, and (3) able to maintain scientific-quality data? 

These questions have driven the science and engineering work in the University of Minnesota’s Real-Time GeoInformation Systems Lab since 2019 and have resulted in a common technology stack with over two thousand sensing devices deployed across three continents. The systems presented here have been moved into a service organization - GEMS Sensing - within the University of Minnesota GEMS Informatics Center to support agricultural sciences. We present our system development as a case study to complement others in the literature \cite{kagan_special_2022} that seek to standardize spatial IoT systems and their use in agricultural science.

In the following sections, we first describe the practical context motivating this work, then the theoretical discussion of upscaling IoT for research and development, followed by the role of open source technologies in upscaling, and finally end with a description of the current IoT system. Throughout, we attempt to emphasize the connection between design principles, scalability, and data integrity and quality to aid future development and deployment of IoT platforms.

\section{Background and Technology Context}
The technologies described in the following sections were established in 2018 through a collaboration between the University of Minnesota (Runck) and BWCS Corp (Morris). Following a lean startup approach, which emphasizes value-creation and technology fit to targeted end-users \cite{osterwalder_business_2010}, open-ended interviews were conducted through 2018 and 2019 with roughly fifty agricultural scientists spanning universities, government, non-profits, and the private sector positions. As interviews progressed, topics began to focus on scientists’ needs related to agro-environmental sensing.

The four major points resulting from these interviews were:
\begin{enumerate}
        \item Scientists had good sensors and loggers in the field, but those were too expensive to achieve high spatial densities.
        \item Scientists had access to large pools of computational resources through either on-site supercomputers or cloud-based technologies.
        \item Scientists had business and agriculturally relevant questions they could not answer because they were not able to collect sufficient data at the right price point to justify to business managers or funders.
        \item In cases where scientists had sufficient funds to broadly deploy sensors, they struggled to maintain data quality across large, open source deployments and connect the data streams to available computing resources in a timely manner.
\end{enumerate}

In sum, sensors and loggers were perceived to be expensive and hard to maintain, and if this challenge could be addressed for scientists, then they would use them more intensively. Considering these observations, we prototyped multiple end-to-end IoT systems and tested them in varied contexts over the course of four years (Table 1). In the remainder of the paper we elaborate on the overall need for upscaling and describe salient decisions surrounding the versioning of the system.

\subsection{The Need to Up-Scale Spatial IoT for Research and Development}

Upscaling IoT applications beyond the research and development stage is a critical area of research for two main reasons. First, to ensure machine learning models scale beyond the research context in agriculture, IoT technologies need to be deployed at a scale where they can collect data of sufficient spatial coverage so that the subsequent models built can generalize across the spatial extent of an entire production system \cite{xie_research_2023}. In technical terms, IoT systems need to collect training and test datasets for the entire range of the potential outcomes in a target agricultural system \cite{runck_state_2023}. 

The validity of current AI and ML systems most often requires that application areas are not “out of sample”, whereby part of the outcome space is left unobserved. Furthermore, differences exist between controlled, on-station experimental settings and on-farm settings, where data may be substantially noisier than monitored and controlled experiments. Thus, data collected at multiple relevant locations, both on-station and on-farm, provide models with the best chance of observing the full range of potential input-output relations. This reason alone - the need to ensure model validity in the target domain - warrants scaling IoT systems.

Even though the ability to scale up IoT technologies impacts scientific generalizability and commercialization potential, it creates unique engineering and logistical challenges. In other words, researchers generally lack the incentives to test applications at scale because it often isn’t necessary for publication – the primary objective of most academic researchers - and the costs of engineering scalable systems are high if a well-established, standardized technology stack does not exist. 

These challenges can be understood in terms of technology readiness levels \cite{noauthor_technology_2023}. In this framework, agricultural scientists lack cost-effective and transparent research platforms to advance spatial IoT technologies beyond TRL 5 in the academic context (Figure 1). Considered considering the interviews described above, the cost of loggers represents a challenge at TRL 5 (field validation of lower cost devices) and the challenges of managing distributed fleets is at TRL 8. Advancing end-to-end spatial IoT technologies from TRL 5 to TRL 8 thus represents the research challenge. 

We term this general challenge of scaling from TRL 5 through 8 the Agricultural IoT Scaling Gap, which we believe hinders the broad adoption of technologies by domain scientists and industry (Figure 1). We repeatedly heard this challenge as an area of concern for plant breeders, agronomists, and other non-digital technology focused researchers, and see addressing this gap as a foundational challenge in applied IoT research and development within digital agriculture.

\begin{figure} % picture
  \centering
  \includegraphics[width=\linewidth]{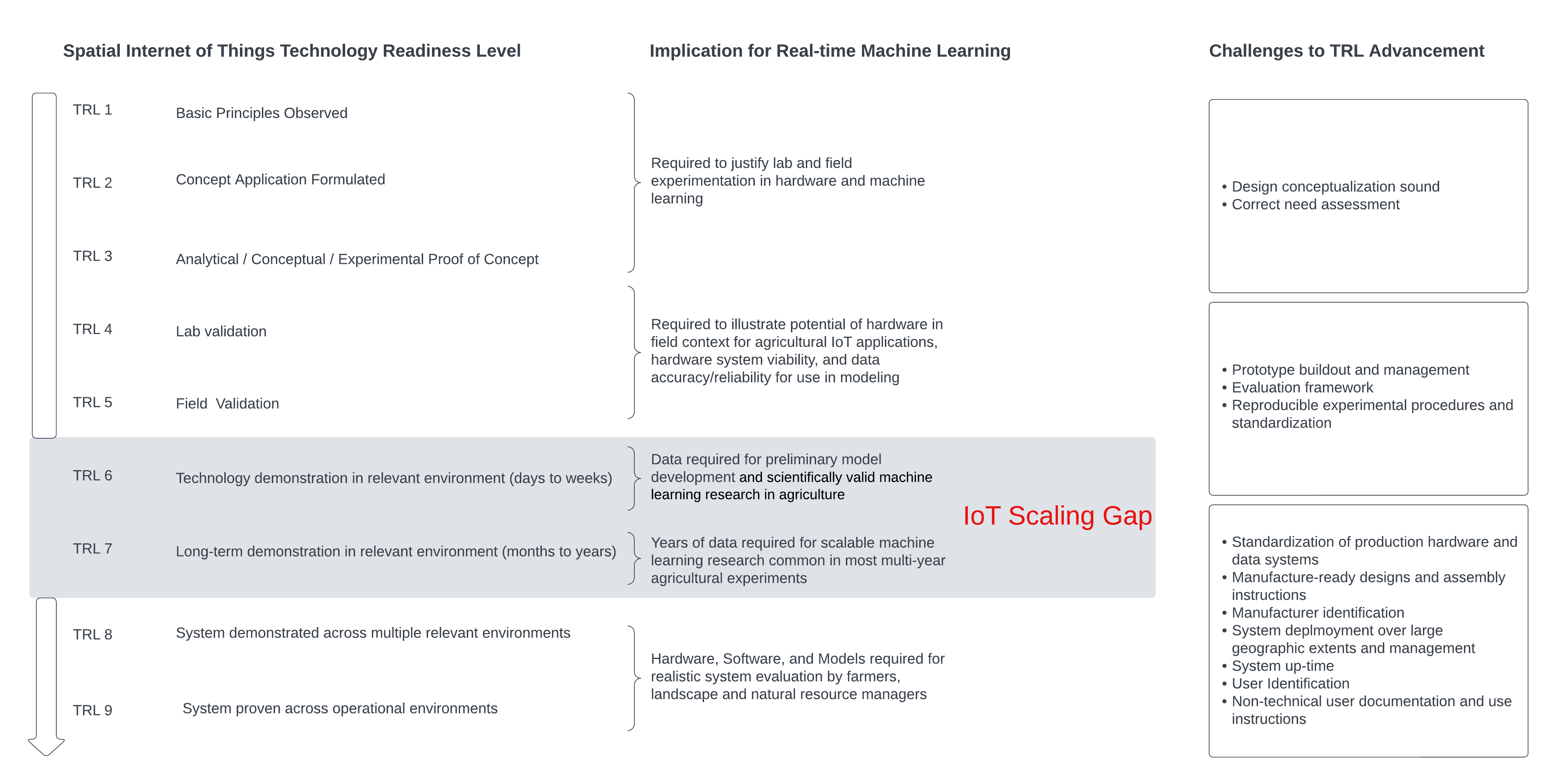}
  \caption{The IoT Scaling Gap prevents research and development in real-time systems from getting to the scale of deployments necessary for broader impacts on agricultural science and productization.}
  \label{fig:fig1}
\end{figure}

\subsection{The Role of Open Source in UpScaling IoT}

The agricultural IoT scaling gap is a general problem, but is pervasive in open source technology development. For this reason, we focus our discussion on this sub-area of IoT development.

One aspect of this scaling gap is related to the open source development environments that are used in agricultural research and development in IoT; for example, there are over 150 publications in the pages of this journal alone relying on open source hardware. Even though progress continues to be made in this space in the not-for-profit and for-profit sectors (see Particle, Conservify, SparkFun, Northern Widget, among others), there remains considerable opportunity to further systematize and expand the general domain for IoT applications to ensure scalability of hardware and firmware, as well as manufacturing pipelines and software systems for managing IoT deployments. While many commercial IoT companies exist (e.g. METER, Arable), the research and development community continues to use open source solutions because of their accessibility, transparency, low cost, and flexibility. This makes open source an essential part of upscaling IoT for agricultural scientists.

In principle, open source technologies offer many advantages for spatial IoT in agriculture. Such systems have licensing agreements (e.g. MIT, GPL, CC BY) that facilitate sharing design files and code publicly with unrestricted or semi-restricted use. This can lead to larger collective benefits as modularized technologies can be easily adopted and applied across a wider scope of application areas \cite{langlois_hackers_2008}. Open source systems are intended to be more in line with the scientific objectives of replicability, reproducibility, and transparency, and the public goods-oriented objectives of transparency and egalitarianism of many scientists. In the case of spatial IoT, where the development stack cuts across multiple engineering disciplines focused on hardware, firmware, backend, and frontend software, open source modules are a thriving part of practice. Yet, translating systems from the benchtop to dozens or hundreds of devices (TRL 5 to TRL 8) remains a challenge for the scientific community, particularly those not specifically trained in digital science and engineering \cite{mao_low-cost_2019}. 

The need for upscaling and the challenges facing open source technology motivate this work. In the following, we report on the development trajectory of a spatial IoT system in the RTGS Lab since 2018 to develop an open source set of modules and turnkey services for researchers. To further support scaling and use by researchers, we have created a service organization in the University of Minnesota called GEMS Sensing (https://gems.umn.edu/services/gems-sensing). We describe the functional specification, evaluation framework, design principles, and system implementation details. Then, we report on the use of the system in four application contexts before concluding with discussion of opportunities for further work on upscaling open technologies to hasten the transfer of digital agricultural technologies beyond research and development contexts.

\section{System Development and Description}

Drawing on the pragmatic and theoretical considerations described above, we proceeded through 14 design iterations of end-to-end IoT systems. This work resulted in the testing and deployment of over 337 nodes and 2,727 sensing devices. 

Nodes have been deployed across the United States (Minnesota, Iowa, Idaho, Oregon, Illinois, Indiana, South Dakota, Massachusetts, Washington, Wisconsin, Michigan, New York, North Carolina, Florida, Hawaii), southern Canada (Ontario, Manitoba), Norway, Finland, Denmark, South Africa, and Malawi. Each set of deployments came with system upgrades and lessons learned which are summarized in Table 1. In the following, we provide a high level summary of a subset of these details.

\begin{longtable}
{|>{\hspace{0pt}}p{0.065\linewidth}|>{\hspace{0pt}}p{0.052\linewidth}|>{\hspace{0pt}}p{0.456\linewidth}|>{\hspace{0pt}}p{0.171\linewidth}|>{\hspace{0pt}}p{0.19\linewidth}|}

\caption*{Table 1. Multiple prototypes of the spatial IoT system were built and tested. The current system is at Version 3.0 and is now governed by the GEMS Sensing Service Organization.}\\ \hline

\multicolumn{1}{|>{\hspace{0pt}}m{0.09\linewidth}|}{Version and Name} & \multicolumn{1}{>{\hspace{0pt}}m{0.052\linewidth}|}{Fleet Size \par{}Sensor Count\par{}(Year)} & \multicolumn{1}{>{\hspace{0pt}}m{0.456\linewidth}|}{IoT Technology Stack} & \multicolumn{1}{>{\hspace{0pt}}m{0.171\linewidth}|}{Applications} & \multicolumn{1}{>{\hspace{0pt}}m{0.19\linewidth}|}{Lessons Learned with Users} \endfirsthead \hline
\vcell{0.1.0 Hope} & \vcell{3 nodes plus gateway \par{}6 sensing devices\par{}(2018)} & \vcell{Communications: LoRaWAN, LoRa32u4, RaspberryPi Base station\par{}Power Supply: Battery\par{}On logger diagnostics: None\par{}External Sensors: humidity and temperature (DHT20) x2\par{}Database: MongoDB\par{}Compute Infrastructure: Atlas DB for MongoDB\par{}Visualization: Kepler.GL\par{}Manufacturing: BWCS Corp lab\par{}Primary users: developers of system} & \vcell{Mobile sensing of microclimate} & \vcell{1. Cost effective per node (\textasciitilde{}\$30)\par\null\par{}2. Good range with line of site\par\null\par{}3. Limited sensor functionality} \\*[-\rowheight]
\printcelltop & \printcelltop & \printcelltop & \printcelltop & \printcelltop \\* \hline
\vcell{0.1.1 Halfway with Photon Gateway} & \vcell{50 nodes plus gateway\par{}300 sensing devices\par{}(2019)} & \vcell{Logger and Communications:\par{}LoRa, LoRa32u4, Particle Photon backhaul via WiFi\par{}Power Supply: Battery and solar\par{}On logger diagnostics: None\par{}External Sensors: Below canopy air temperature, humidity, barometric pressure (Bosch BME280), soil temperature (Maxim DS18B20), soil moisture (DF Robot SEN0193 Capacitive Soil Moisture Sensor v1.0\footnote{ \url{https://wiki.dfrobot.com/Capacitive_Soil_Moisture_Sensor_SKU_SEN0193}}\par{}), photosynthetically active radiation (TCS230 Color Recognition Module with custom calibration)\par{}Database: MongoDB\par{}Compute Infrastructure: Atlas DB for MongoDB; Google Cloud Platform for all else\par{}Software: Particle.io webhooks, NodeJS webserver and API\par{}Visualization: HTML, CSS, Jupyter Notebooks\par{}Manufacturing: Marchetto Lab\par{}Primary users: developers of system} & \vcell{In-field below canopy microclimate monitoring} & \multirow{8}{\linewidth}{\hspace{0pt}
\begin{enumerate}[wide, labelwidth=!, labelindent=0pt]
	\item Particle platform rate limited data packet throughput
	\item Cost of nodes affordable \textasciitilde{}\$135 each
	\item Substantial packet loss or corruption for nodes without line of sight to gateway
	\item Setup of nodes was labor intensive requiring skilled technicians
	\item Sensor calibration was time intensive and could only included side-by-side point samples
	\item Management of nodes was labor intensive because of power management
	\item Node firmware fixes were labor intensive because firmware could only be updated via USB
	\item Manufacturing was done in-Lab and suffered from variable quality and large amounts of labor
\end{enumerate}
\vspace{100pt}
} \\

\printcelltop & \printcelltop & \printcelltop & \printcelltop &  \\* \cline{1-4}
\vcell{0.1.2 Halfway with Open Source Gateway} & \vcell{Same as 0.1.1} & \vcell{Same as 0.1.1 unless noted\par{}Communication: LoRa, LoRa32u4, RaspberryPi gateway with tailing agent implemented in Python sent via WiFi\par{}Database: On RaspberryPi, SQLite was used for data storage before confirmation of checks being sent} & \vcell{Same as 0.1} &  \\*[-\rowheight]
\printcelltop & \printcelltop & \printcelltop & \printcelltop &  \\* \cline{1-4}
\vcell{0.1.3 Halfway Barley Winter Kill} & \vcell{Same as 0.1.1} & \vcell{Same as 0.1.1} & \vcell{Environmental drivers of winter kill in Winter Barley} &  \\*[-\rowheight]
\printcelltop & \printcelltop & \printcelltop & \printcelltop &  \\ \hline
\vcell{0.2.0 Compost Monitoring} & \vcell{2 nodes\par{}16 sensing devices\par{}(2019)} & \vcell{Same as 0.1.1 unless noted\par{}External Sensors: temperature x 4 (Maxim DS18B20) and solar radiation x 4 (NOYITO TCS230 Color Recognition Module with custom calibration)\par{}Software: Flask API\par{}User interface: Email notifications\par{}Manufacturing: in Marchetto Lab\par{}Primary users: developers of system and single field manager} & \vcell{Compost temperature monitoring} & \vcell{1. LoRa connectivity around buildings problematic due to line of sight requirement} \\*[-\rowheight]
\printcelltop & \printcelltop & \printcelltop & \printcelltop & \printcelltop \\ \hline
\vcell{1.0.0 Palmer LoRa} & \vcell{6 nodes\par{}54 sensing devices\par{}(2019-2020)} & \vcell{Logger and Communications:\par{}two systems were used: LoRa, LoRa32u4, RaspberryPi gateway with tailing agent implemented in Python\par{}Power Supply: DC Bus (5V) or 24V AC\par{}On logger diagnostics: Power Source Sensing\par{}External Sensors: Soil moisture x3 (AITRIP Capacitive Soil Moisture Sensor v1.1), temperature x 4 (Maxim DS18B20), CO2 (Sensirion SGP30 ), O2 (Amphenol SGX-4OX ), solar radiation (AMS TCS3400)\par{}Database: MongoDB; on RaspberryPi, SQLite was used for data storage before confirmation of checks being sent\par{}Compute Infrastructure: Atlas DB for MongoDB; Google Cloud Platform for all else\par{}Software: Flask web server and API\par{}Visualization: HTML, CSS, Jupyter Notebooks\par{}Manufacturing: in RTGS Lab\par{}Primary users: developers of system and the Watkins turfgrass breeding program} & \vcell{Characterize microclimate and gas exchange on golf course greens} & \vcell{1. \$185 per node (no labor)\footnote{Remote unit = $30 + $9 + $50 + $5 = $94Main unit = $20 + $20 + $15 = $55Sensor = $2x4 + $(2 + 2)*3 = $20Power Supply = $15Total = ~$185 (No labor)}\par\null\par{}2. Local logging of data essential with spotty connectivity for LoRa.\par\null\par{}3. Addressed most power management problems in version 0} \\*[-\rowheight]
\printcelltop & \printcelltop & \printcelltop & \printcelltop & \printcelltop \\ \hline
\vcell{2.0.0 Palmer Cellular} & \vcell{2 nodes\par{}18 sensing devices\par{}(2019-2020)} & \vcell{Logger and Communications: GEMS Palmer logger, Particle Boron Cellular\par{}Power supply: Battery and solar\par{}On logger diagnostics: Power Source Sensing\par{}External Sensors: Same as 1.0.0\par{}Database: MongoDB\par{}Compute Infrastructure: Atlas DB for MongoDB; Google Cloud Platform for all else\par{}Software: Flask web server and API. Particle Boron, Particle.io webhooks\par{}Visualization: HTML, CSS, Jupyter Notebooks\par{}Manufacturing: in RTGS Lab\par{}Primary users: developers of system and the Watkins turfgrass breeding program} & \vcell{Characterize microclimate and gas exchange on golf course greens} & \vcell{1. Cost of node increased roughly \$60 over version 2.0.0 Palmer LoRa\par\null\par{}2. Cellular more robust and more reliable packet transmission over large areas.\par\null\par{}3. Cellular eliminates the need for base station site selection, increasing installation speed and ease.} \\*[-\rowheight]
\printcelltop & \printcelltop & \printcelltop & \printcelltop & \printcelltop \\ \hline
\vcell{2.0.1 Irrigation Particle Boron} & \vcell{16 nodes\par{}64 sensing devices\par{}(2019 - still in operation)} & \vcell{Logger and Communications: GEMS Hyena logger, Particle Boron Cellular\par{}Power supply: Battery and solar\par{}On logger diagnostics: Temperature, Humidity, Sensor bus overcurrent, SD overcurrent, Ambient light sensing\par{}External Sensors: soil moisture x 3 (Vegetronix VH400), soil temperature x 1 (Maxim DS18B20)\par{}Database: MongoDB\par{}Compute Infrastructure: Atlas DB for MongoDB; Google Cloud Platform for all else\par{}Software: Flask web server and API. Particle Boron, Particle.io webhooks\par{}Visualization: Jupyter Notebooks\par{}Manufacturing: in RTGS Lab\par{}Primary users: developers of system and the Sharma lab} & \vcell{Irrigation scheduling} & \vcell{1. Unstructured data in MongoDB resulting in slow query times and high costs of data wrangling post-hoc\par\null\par{}2. Manufacturing of greater than 10 nodes unrealistic for high levels of quality assurance and control\par\null\par{}3. Need improvements to firmware for low light parts of season under canopy in order to ensure continuous operation on battery charge only} \\*[-\rowheight]
\printcelltop & \printcelltop & \printcelltop & \printcelltop & \printcelltop \\ \hline
\vcell{2.0.2 Malawi MESONET} & \vcell{80 nodes\par{}640 sensing devices\par{}(2021 - still in operation)} & \vcell{Logger and Communications: Particle Boron Cellular\par{}Power supply: Battery and solar (custom charge controller)\par{}On logger diagnostics: Same as 2.0.1, plus on board temperature and humidity\par{}External Sensors: rain gauge (Misol tipping bucket), soil moisture sensor (Vegetronix VH400), soil temperature sensor (Maxim DS18B20), air temperature/humidity/pressure (GEMS Haar Primal v0.0), anemometer (Modern Device Wind Sensor Rev P), solar radiation (AMS TCS3400)\par{}Database: PostGIS\par{}Compute Infrastructure: Google Cloud Platform\par{}Software: Flask web server and API behind NGINX and gunicorn. Particle Boron, Particle.io webhooks\par{}Visualization: Jupyter Notebooks, REACT\par{}Manufacturing: XOR in Capetown, South Africa\par{}Primary users: Broad-based stakeholders across the country including public and private sectors} & \vcell{Meso-scale real-time weather monitoring} & \vcell{1. Implementation of Data Flow from Raw to Level 1. Switch to PostGIS with a raw table retains flexibility of MongoDB with query time benefits of structured database\par\null\par{}2.External manufacturer addressed the challenge of quality assurance and control of hardware, but doing so internationally generated substantially logistical challenges around shipping exacerbated by pandemic\par\null\par{}3. Version 2 systems too technical for most field technicians to install and operate even with detailed technical videos and instructions} \\*[-\rowheight]
\printcelltop & \printcelltop & \printcelltop & \printcelltop & \printcelltop \\ \hline
\vcell{2.0.3. Minnesota MESONET Open Source} & \vcell{11 nodes\par{}99 sensing devices\par{}(2021 - still in operation)} & \vcell{Logger and Communications: Particle Boron Cellular\par{}Power supply: Battery and solar (custom charge controller)\par{}On logger diagnostics: Same as 2.0.2\par{}External Sensors: Rain gauge (Davis AeroCone), air temperature/humidity/pressure (GEMS Haar Primal v2.0), soil moisture (METER TEROS 10), soil temperature (Maxim DS18B20), solar radiation (Apogee SP-212-SS), anemometer (InSpeed version II hall effect), wind direction (InSpeed e-Vane II)\par{}Database: PostGIS\par{}Compute Infrastructure: Google Cloud Platform\par{}Software: Flask webserver and API behind NGINX and gunicorn. Particle Boron, Particle.io webhooks\par{}Visualization: Jupyter Notebooks, REACT, iFrame visualization on Research and Outreach Center websites\par{}Manufacturing: in RTGS Lab\par{}Primary users: Research Faculty and Staff; local growers and community members} & \vcell{Meso-scale real-time weather monitoring for low cost and high density monitoring} & \vcell{1. Version 2 systems too technical for most field technicians to install and operate even with detailed technical videos and instructions\par\null\par{}2. Data visualization required for less technical users requires intensive user interface iteration and new technical operating procedures\par\null\par{}3. Standardization of naming conventions across system types and deployment configurations becomes a challenge because they each require different quality assurance and control procedures in the database.} \\*[-\rowheight]
\printcelltop & \printcelltop & \printcelltop & \printcelltop & \printcelltop \\ \hline
\vcell{2.0.4. Minnesota MESONET Campbell} & \vcell{6 nodes\par{}54 sensing devices\par{}(2021 - still in operation)} & \vcell{Logger and Communications: Campbell Scientific CR1000X-NA-ST-SW\par{}Power supply: Battery and solar\par{}On logger diagnostics: None\par{}External Sensors: rain gauge (Texas Electronics TE525WS-L15-PT), wind speed/direction (RM Young 05103-L15-PT), temperature/relative humidity (CSL Digital Temperature/RH Sensor HygroVUE10-17-PT), solar radiation (Apogee Digital Thermopile pyranometer CS320-10-PT), barometric pressure (Setra 278 Barometer CS100), soil moisture/temperature (CSL Water Content Reflectometer Plus CS655-17-PT-DS)\par{}Database: PostGIS\par{}Compute Infrastructure: Google Cloud Platform\par{}Software: Flask webserver and API behind NGINX and gunicorn. Campbell LoggerNet.\par{}Visualization: Jupyter Notebooks, REACT\par{}Manufacturing: none\par{}Primary users: Research Faculty and Staff; local growers and community members} & \vcell{Meso-scale real-time weather monitoring to establish standards for cross-calibration} & \vcell{1. Data user trust in brand names prominent. Side-by-side comparison of data on an on-going basis increases trust in open source systems\par\null\par{}2. Data visualization required for less technical users requires intensive user interface iteration and new technical operating structures\par\null\par{}3. Standardization of naming conventions across system types and deployment configurations becomes a challenge because they each require different quality assurance and control procedures in the database.} \\*[-\rowheight]
\printcelltop & \printcelltop & \printcelltop & \printcelltop & \printcelltop \\* \hline
\vcell{2.0.5.A Palmer Boron Cellular} & \vcell{31 nodes\par\null\par{}651 sensing devices\par\null\par{}(2021 - still in operation)} & \vcell{Logger and Communications: Particle Boron Cellular\par{}Power supply: Battery and solar (custom charge controller)\par{}On logger diagnostics: Same as 2.0.2\par{}External Sensors: soil moisture x 3 (Vegetronix VH400), soil temperature x 3 (Maxim DS18B20), carbon dioxide sensor x1 (GEMS Hedorah v0.0), oxygen sensor x1 (Apogee SO-210), solar radiation (AMS TCS3400)\par{}Database: PostGIS\par{}Compute Infrastructure: Google Cloud Platform\par{}Software: Flask webserver and API behind NGINX and gunicorn. Particle Boron, Particle.io webhooks\par{}Visualization: Jupyter Notebooks, REACT\par{}Manufacturing: in RTGS Lab\par{}Primary users: Large team of researchers from across multiple research institutions; visualization for golf course superintendents and extension} & \vcell{Characterize microclimate and gas exchange on golf course greens and fairways} & 
\multirow{2}{=}{
\vspace{-100pt}
\begin{enumerate}[wide, labelwidth=!, labelindent=0pt]
\item Nodes performed well under two inches of ice. 
\item Able to be setup by golf course superintendents across northern United States and Canada 
\item Form factor not specific enough for golf course superintendents, they would prefer it fit in cup hole or be off the surface of the green 
\item Need to standardize logger and sensors supported by our system to enable more consistent service delivery to large research groups 
\item Need to increase the ease of manufacturing
\end{enumerate}
} \\*[-\rowheight]
\printcelltop & \printcelltop & \printcelltop & \printcelltop &  \\* \cline{1-4}
\vcell{2.0.5.B Palmer Boron Cellular} & \vcell{13 nodes\par{}273 sensing devices\par{}(2021 - still in operation)} & \vcell{Same as 2.0.5.A except where noted\par{}External Sensors: carbon dioxide sensor x2 (GEMS Hedorah v0.0), oxygen sensor x2 (Apogee SO-210), solar radiation (AMS TCS3400)} & \vcell{} &  \\*[-\rowheight]
\printcelltop & \printcelltop & \printcelltop & \printcelltop &  \\ \hline
\vcell{3.0.0 Flight} & \vcell{117 nodes\par{}570 sensing devices\par{}(2022 - on-going)} & \vcell{Logger and Communications: GEMS Kestrel logger, Particle B Series SoM\par{}Power supply: Battery and solar\par{}On logger diagnostics\par{}Standardized External Sensors across Projects: soil electrical conductivity, relative dielectric permittivity, volumetric water content, soil temperature (Acclima TDR-315 series), oxygen (Apogee SO-421 and AO-001), solar radiation (Apogee SP-212 and SP-421-SS), rainfall (Davis AeroCone), air temperature and relative humidity (Haar: Sensirion SHT31), barometric pressure (Haar: Infineon, Base DPS368XTSA1), carbon dioxide, humidity, temperature, VOC and Carbon Dioxide equivalent (Sensirion SCD30), wind speed (InSpeed Version II Hall Sensor), wind direction (InSpeed E-Vane II)\par{}Databases: PostGIS\par{}Compute Infrastructure: Google Cloud Platform; Minnesota Supercomputing Institute\par{}Software: Flask webserver and API behind NGINX and gunicorn. Particle Boron, Particle.io webhooks\par{}Visualization: Jupyter Notebooks, REACT, Grafana\par{}Manufacturing: Caltronics Design and Manufacturing, Stacy, Minnesota\par{}Primary users: Multiple for-profit companies, large multi-institution research projects, non-governmental organizations} & \vcell{Multiple application areas including MESONETs, golf course microclimate and gas sensing, irrigation sensing, site-specific monitoring for crop modeling and breeding, pesticide drift risk monitoring} & \vcell{1. Need to shift governance and service model away from lab\par\null\par{}2. Further improve standardization of infrastructure with rest of University of Minnesota\par\null\par{}3. Further develop sustainable service model for reliable and consistent data collection\par\null\par{}4. Move to longer technology development timeline and explicitly manage and support versions\par\null\par{}5. Establish clear contractual basis with users to ensure service organization meets expectations\par\null\par{}6. Improve logistics around ordering and shipping units, specifically, time from order to delivery/deployment\par\null\par{}7. Improve procedures for developing documentation and standard operating procedures that retain generality but ensure specific data collection needs} \\*[-\rowheight]
\printcelltop & \printcelltop & \printcelltop & \printcelltop & \printcelltop \\ \hline
\vcell{Summary Totals} & \vcell{} & \vcell{} & \vcell{} & \vcell{} \\*[-\rowheight]
\printcelltop & \printcelltop & \printcelltop & \printcelltop & \printcelltop \\ \hline
\vcell{14 system designs} & \vcell{337 Nodes\par{}2,727 sensing devices\par\null\par{}2019 - 2022} & \vcell{} & \vcell{} & \vcell{} \\*[-\rowheight]
\printcelltop & \printcelltop & \printcelltop & \printcelltop & \printcelltop \\ \hline
\vcell{} & \vcell{} & \vcell{} & \vcell{} & \vcell{} \\*[-\rowheight]
\printcelltop & \printcelltop & \printcelltop & \printcelltop & \printcelltop \\ \hline
\end{longtable}

\subsection{User and Use Case Descriptions}
From information collected in the interviews described in Section 2 and based on observations over fourteen system design iterations, we found that users required flexible IoT sensing systems so that they could use the same node (logger and telemetry) with multiple sensing devices across different/varying experimental settings. Furthermore, scientific users prefer to understand how measures are being impacted at each stage in the data flow system, from the point of observation through telemetry and storage. This sort of transparency is desirable across multiple information technology contexts and is used primarily for logging causes of system outages. In the scientific case, logs are desirable to understand data flow and transformations that may affect the quality of the findings derived from downstream data.

Transparency into data systems comes with relatively high maintenance costs. Yet, many research programs are often cost constrained; some seek to deploy long term experiments with live streaming of data but lack sufficient on-going funds to support such deployments. Simultaneously, scientists are relatively price-insensitive when writing grants if data are required for a specific experimental outcome. Because there are typically limits on staff availability for field work, scientists require robust sensing systems that are field hardened. Lastly, scientists are often deploying sensing systems in poorly characterized ways at the start of a new research project, where measurement approaches may still be in flux. Then, as the instrument operating procedures become more standardized, the objective of a system is to facilitate rigorous systematization to reduce person-to-person measurement bias. These general observations about agricultural scientists’ unique needs in IoT systems are derived from multiple specific cases (Table 1).

\subsection{Design Principles}
The underlying design philosophy has been “encapsulation, not obfuscation” \cite{raghavan_means_2017}, which means that users should be able to access complex system-level functionality, such as spatiotemporal quality assurance and control procedures or the underlying hardware designs, without necessarily having to understand a system’s inner workings. In this way, system functionality is encapsulated. At the same time, most private companies would say that they achieve this end - and they do - but where they typically fall short is in the obfuscation of system functionality. From a scientific perspective, obfuscation is particularly problematic for reproducible science because data transformations from data conversion, logging, and calibration are opaque. In this way, the design philosophy intends to allow users to progressively move from higher levels of abstraction to more concrete implementation details throughout the entire hardware and software stack.

Encapsulation also undergirds the design of hardware, firmware, and software sub-components to minimize sub-system dependencies. This approach follows best practices of deep abstraction and functionally interpretable interfaces from software engineering \cite{ousterhout_philosophy_2018}. This approach has resulted in a pool of open source loggers, power systems, firmware, sensors, and quality assurance and quality control (QAQC) pipelines that can be mixed and matched in order to achieve a specific project objective. In combination, these modules can be reconfigured to achieve several different objectives. We elaborate on each set of modules in the following system overview section.

\subsection{System Overview}
\subsubsection{General System Architecture and Data Flow}
The general system architecture consists of nodes deployed in a field or lab that stream data via wireless connection through a publish-subscribe system (such as the Particle Internet of Things Platform or ThingsSpeak) to a full stack web application hosted on Google Cloud platform (Figure 2). Nodes can also be “virtual”, in that a correctly structured data packet can be sent from a server, which is an extract-transform-load procedure from another data store \cite{vassiliadis_survey_2009}. In this way, we support interfacing with non-GEMS hardware, such as Campbell Scientific instruments. Each component varies slightly depending on the project but is deployed following this similar architectural approach.

\begin{figure}[htb] % picture
  \centering
  \includegraphics[width=\linewidth]{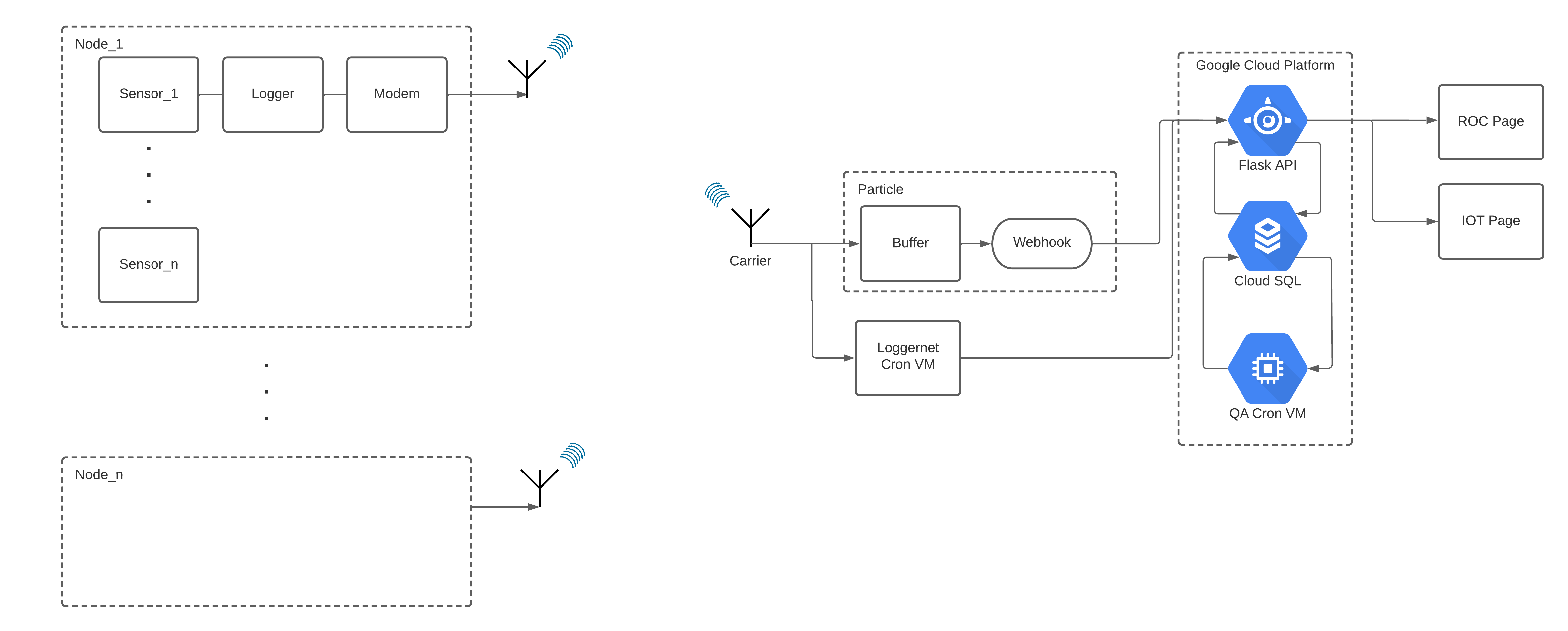}
  \caption{ Illustrates the general system architecture with Particle as the PubSub provider before data are sent to a Flask API for storage and subsequent visualization.}
  \label{fig:fig2}
\end{figure}

\subsubsection{Hardware}
GEMS Sensing open source hardware and firmware modules extend work from the Wickert lab and Northern Widget \cite{wickert_open-source_2019,schulz_development_2021, llc_northern_nodate}. At a high-level, all loggers and sensors provide similar functionality, but in scientific applications, fitting the logger and sensor to the specific problem is necessary to ensure correct data is collected. While to many, every data logger or temperature sensor appears identical and equally applicable across cases, they might have unique targeted applications. In the following, we describe the suite of loggers and sensors supported in the GEMS Sensing ecosystem with description on their target application contexts.

\paragraph{\textit {3.3.2.1 Loggers}}\mbox{} \\
Automated digital data loggers are commonplace across the agricultural and environmental sciences. The following loggers, called Palmer, Hyena and Kestrel, are complementary and grew out of work on Resnik \cite{schulz_development_2021} and Margay \cite{schulz_northernwidget-skunkworksproject-okapi_2021}. Palmer, Hyena, and Kestrel support integration with telemetry, particularly the Particle Internet of Things platform (https://www.particle.io), to support real-time application development in the agricultural and environmental sciences.

\paragraph{\textit {3.3.2.1.1 Palmer}}\mbox{} \\
Engineered originally in 2019, Palmer was developed to fill the gap of low cost, networked data loggers for golf course greens monitoring. Built around the LoRa32u4 module with a minimal dedicated sensor package. The LoRa32u4 consists of an Arduino-based microcontroller and LoRa telemetry system. Via an external port, there was a serial to secure digital card interface module. Because Palmer was developed for a specific application, it lacked modularity and used hardware DC power or 24 volt AC power common in golf course irrigation sprinkler systems.

The gateway for the systems was developed with a LoRa32u4 connected via serial to a Raspberry Pi. A Python script ran onboard as a listener. When a packet was received, it would timestamp the packet, write the packet to an SQLite database, and then read the tail of the database to send packets via WiFi to the webserver. 

The hub and spoke LoRa architecture with nodes and a gateway presented a number of challenges for distributed deployments, namely the need for an appropriately placed gateway that was within line-of-site of the node. Ultimately, these challenges with LoRa led to the shift to cellular-based loggers in future design iterations.

\paragraph{\textit {3.3.2.1.2 Hyena}}\mbox{} \\
Designed in 2020 and updated in 2021, Hyena built on the Palmer design, but with the Particle Cellular Boron at the core. Using the Particle Boron reduced development time because of Particle’s well-integrated technology stack and allowed achievement of real-time telemetry and remote device management mostly off-the-shelf. In support of the Boron, Hyena includes real-time battery-backed clock support, on-board sensing capabilities such as temperature, relative humidity, and ambient light for hardware diagnostics, non-volatile data storage (secure digital card), minimal sensor power management, and a hardware watchdog timer to protect against firmware faults.

While the specific challenges associated with the hub and spoke LoRa architecture were uncovered in the golf course monitoring context, the original motivating use case for Hyena was extremely low cost agrometeorological monitoring in sub-Saharan Africa. The Particle platform was chosen in large part because of the global availability of their syste

\paragraph{\textit {3.3.2.1.3 Kestrel}}\mbox{} \\
Developed from 2021 to 2022, Kestrel is the third iteration growing out of the Palmer design lineage, but deviates in clear ways through enhanced system interchangeability and internal diagnostic monitoring. The Kestrel design sought to maximize flexibility within the sensor ecosystem with a targeted set of devices that communicate via SDI12, and I2C buses. The logger is also designed to work with common non-bus interfaces, specifically analog voltage inputs and pulse based inputs. In this way, while still accommodating the common use cases requested regularly by agro-environmental scientists, the range of potential peripherals is more limited than that of its predecessor. However, this configuration comes with important upsides, not least that the flexibility of use within the ecosystem is dramatically increased. A damaged sensor can be replaced by a novice in a matter of minutes and with little risk of compromising the system operation. These reduced risks stem from two main reasons. First, a damaged sensor is electrically isolated automatically to prevent systemic failure of the system. Second, there is no longer any need for users to open the logger box. This is important because every time the internal system is exposed to the environment, the risk for failure increases along with the likelihood for costly maintenance. 

At the high level, Kestrel is built around the Particle B404 cellular module, which is a logical continuation of the Particle Boron, but more suitable for manufacturing. The core organizing principle for peripherals interfaces are daughterboards which in this ecosystem are dubbed talons. Talons provide the electrical interface to, and monitoring circuitry for, sensors in the system. In this architecture, the main logger board handles the core logger functionality - timekeeping, internal sensors, data storage, user interface, power management, etc. The main logger board then has a set of generalized inputs, each one of which is designed to connect to a talon. These talon daughterboards then provide the specific electrical interfaces required for different sensors. This allows the system to be easily adjusted by changing the set of talons used to vary the potential sensor loadout of the system. Additionally, improvements in usability were made in the mechanical design of the system, including an external power switch, external charging and computer interface, and polarized circular connectors. These external connectors facilitate the desired flexibility and the design of the talons is based around a principal of isolation - which is key to the system robustness.

\paragraph{\textit {3.3.2.2 Procurement, Manufacturing, Shipping, and Sensor Calibration}}\mbox{} \\
Before 2021, all hardware was manufactured in labs at the University of Minnesota with only PCB manufacture done externally. As the number of devices required for deployments has increased, the need emerged to identify external manufacturers.

For deployments in sub-Saharan Africa with the Hyena logger, XOR Manufacturing based in Cape Town, South Africa was used. Parts were procured directly by XOR in some cases, but in others University of Minnesota procured parts from vendors and re-boxed them for shipment to XOR. This was necessary due to limited availability of certain components in South Africa. Working in an international context brought unexpected logistical challenges from coordinating with international shippers to navigating complex customs and import tax requirements.

For deployments in the United States, Canada, and Europe, we contracted with Caltronics Design and Assembly of Stacy, Minnesota. An ISO 9001 certified manufacturer based nearby the University of Minnesota, Caltronics handled all the procurement, assembly, and bulk shipment of units back to the University of Minnesota. Bulk shipments were then re-boxed with additional documentation and then shipped out to their final location for installation by University of Minnesota staff.

For seasonally-deployed projects, devices were shipped back to the University of Minnesota for inspection, maintenance, and random spot checks for sensor drift. Perennially deployed devices were serviced by local technicians with support by the RTGS Lab.

\paragraph{\textit {3.3.3 Firmware and Logging Intervals}}\mbox{} \\
Both LoRa32u4 and Particle devices are Arduino compatible systems. This means that code written for an Arduino is most often compatible with these devices. Across the firmware, logging and telemetry intervals have been set for the specific application and are tuned to the requirements of each sensor. For example, pulse-based sensors (e.g. cup anemometers or rain gauges) require on-going counts to be recorded, whereas other sensors can be turned on (and off) to sample only once per recording interval (e.g. gas sensors such as CO2 and O2). 

Across our application areas, logging has happened at everywhere between 5 minute, 15 minute, and hourly intervals. Logs are flushed to non-volatile memory every 15 minutes with hourly backhaul via telemetry. Regardless of telemetry failure or success, data is stored locally until memory is full.

All of the Particle firmware has been written in C++14 and is designed using a modular, class-based architecture. In firmware for Kestrel, each sensor is a subclass of a general sensing device class that specifies standard methods and interfaces exposed to the main logger program. In addition to sensor classes, other devices - such as the real-time clock, SD card, modem, and breakout board (daughter board) - also are abstracted as modules. Lastly, the logger itself orchestrates the submodules and ensures overall functioning of the system.

A major challenge for the firmware was battery management in low- and no-solar input conditions. The main power draw comes from the cellular modem on the Particle Boron. With the modem running, power draw is approximately 100 mA during wake cycles and between 10 mA and 20 mA during Particle’s default sleep operations. With the modem off, power draw is roughly 6 mA awake and 1.2 mA asleep. For this reason, in cases of low and no-solar, the firmware manages the modem by powering it off except for one time per hour when data from the last four readings is sent back to central servers. In cases where solar is not a limiting factor, the modem switches to a reduced power setting between logging events, while remaining connected to the network. This balances the need for reasonable power consumption with the desire for active response from the system.

\begin{figure}[htb] % picture
  \centering
  \includegraphics[width=\linewidth]{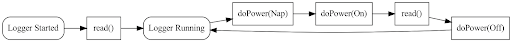}
  \caption{Illustration of the firmware routine for sensor reading and power management via modules.}
  \label{fig:fig3}
\end{figure}

\paragraph{\textit {3.3.4	Sensors and Communication Interfaces}}\mbox{} \\
Throughout the lifespan of development, over 65 sensing devices have been considered, developed, and tested. Throughout all projects, sensors constrained both the design of data loggers and software systems. Version 3 systems support I2C, SDI12, and analog communication and a standard set of sensors that including Acclima TDR-315H soil moisture and temperature sensors, Apogee 212-SS Pyranometer, Apogee SO-421 Oxygen gas concentration, Bosh BME 280, Davis Aerocone 6466M, InSpeed E-Vane II and Version II Hall Sensor, and the Sensirion SCD30.

\paragraph{\textit {3.3.5	Application Programming Interfaces, User Interfaces, and Data Sharing}}\mbox{} \\
The GEMS Sensing user interface evolved from simple HTML and CSS pages (Figure 4A) into a REACT.JS frontend (Figure 4D) and then into a Grafana-based set of organizations and dashboards (Figure 4G). Surveys were sent to 78 users of the Grafana v3 dashboard to evaluate experiences with a 35.89\% response rate (n=28). One response was removed because the person stated in the open ended comments that they did not use the tool, but filled out the survey anyway. User responses indicated that the dashboard were visually appealing (>89\% appealing or very appealing) sufficiently easy to navigate (>82\% responded easy or very easy), provided a positive user experience (>91\% satisfied or very satisfied), and trusted the accuracy of the data (>86\% trust or strongly trust).

\begin{figure}[htb!]
    \centering
    \includegraphics[width=.9\textwidth]{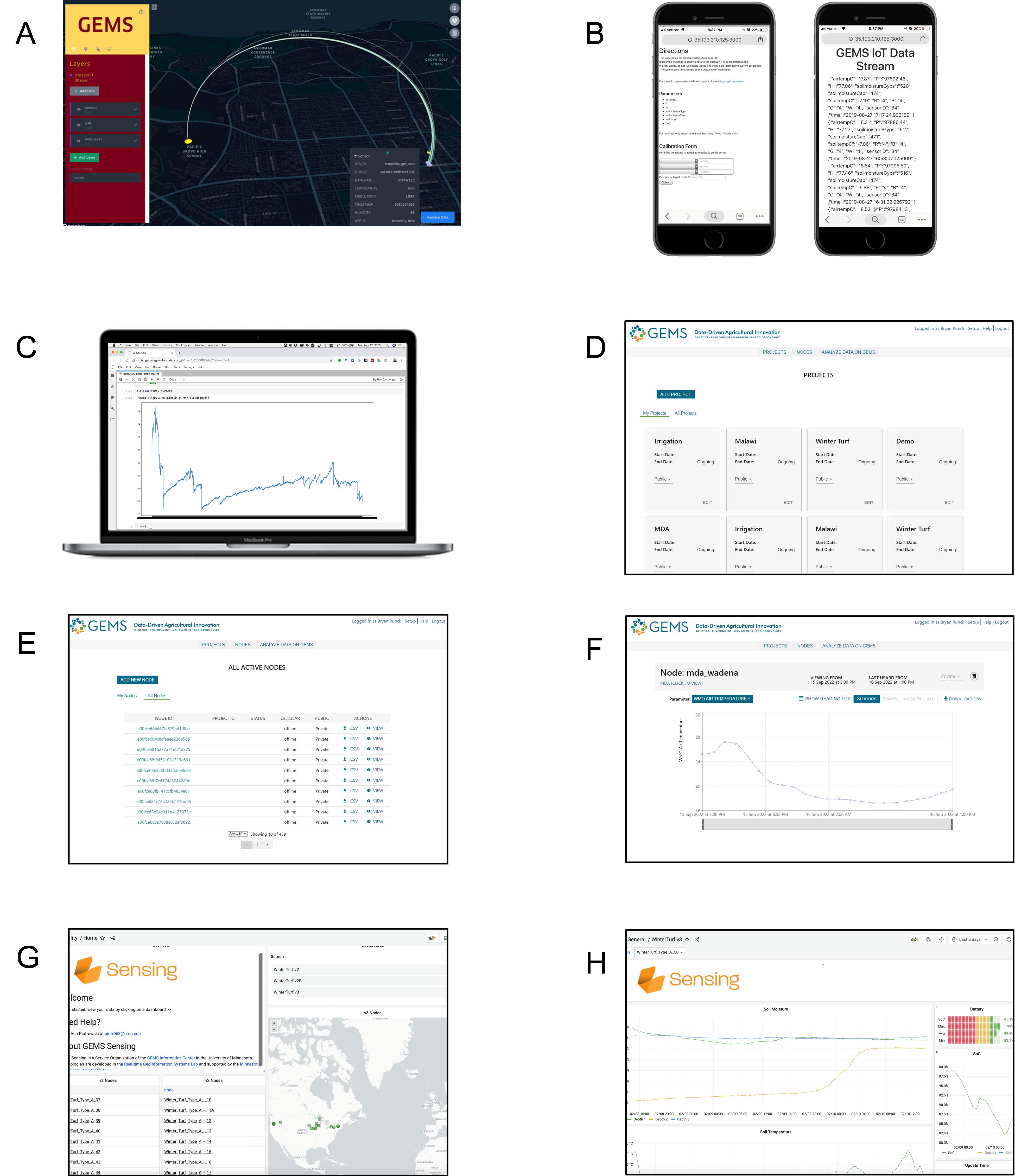}
    \caption{Examples of user interfaces through time. A) v0.1 User Interface built using Keplar.gl shows the last packet from each node. B) Shows the v0.1 series mobile website for side-by-side calibration data entry and the continuous fleet data stream. C) The v0.1 visualization of data in a jupyter notebook accessible via NodeJS application programming interface. D) the v2 system project page allowing users to both share and download data from all of the nodes in a project. E) The v2 System list of nodes available to a user. F) The v2 System time series visualization from a Campbell Scientific node. G) Grafana v3 dashboard for fleetwide node summary. H) Grafana v3 dashboard for single node.}\label{fig:4}
\end{figure}

\paragraph{\textit {3.3.6	Standard Operating Procedures and User Documentation Support}}\mbox{} \\
Project sensing requirements are either already standardized, as in the case of meteorological air temperature, or are unstandardized, as is the case with novel phenotype characterization. Deployment-specific operating procedures are developed with any applicable standard documentation and linked to each individual device utilizing standardized field names or QR codes and via the data dashboard. 

This approach addresses the ambiguity that can arise from the same device being used across different operating procedures resulting in non-comparable measures. Each standard operation procedure (SOP) is developed uniquely for a project, but builds on standard modules when the same parameter is being sensed. For each project, we link all SOPs and helpful links within the user interface’s Grafana-based homepage. For non-standardized measures, procedures are developed in coordination with domain specialists. In addition to standard operating procedures, help videos and a knowledge base are under on-going development to respond to repeated and standard questions for large fleets.

\paragraph{\textit {3.3.7	Data Quality Assurance and Control Procedures}}\mbox{} \\
User standards and data quality labeling are utilized to support managing GEMS Sensing data flow systems as they advance from small lab and field experiments with devices (TRL 4 and 5) to managing data for larger deployments (TRL 7-8) (Figure 5). Specifically, data management is conceptualized as a flow from raw to Level 2 data, where each level involves an increase in the quality that the system will provide. The performance standards for each step of the data flow are described in Figure 2 and encapsulate the complexity of QAQC for users. The approach allows users to access data at all stages for interrogation. If users provide specific parameters recognized by the system with data packets referenced in space and time, software can automatically clean, process, and visualize the data. Alternatively, in cases where users are developing systems in an iterative, ad hoc manner, they can still use modules within the IoT platform, but with correspondingly less processing applied. The goal of this approach is to both allow for rapid, informal prototyping - as is common across most electronics labs (TRL 4) - while also supporting rigorous quality control and assurance procedures at scale (TRL 7 and 8). 

For version 3.0 of the system used as the basis for services to researchers, Level 0 data processing is done at the time of collection and database storage because it can be standardized across all project types.  Level 1 processing (missing and null value checks, hardware range checks, and user-specified outlier detection) is done when the data are queried from the database before visualization in Grafana.

\begin{figure}[hbt!] % picture
  \centering
  \includegraphics[width=\textwidth,height=0.4\textheight,keepaspectratio]{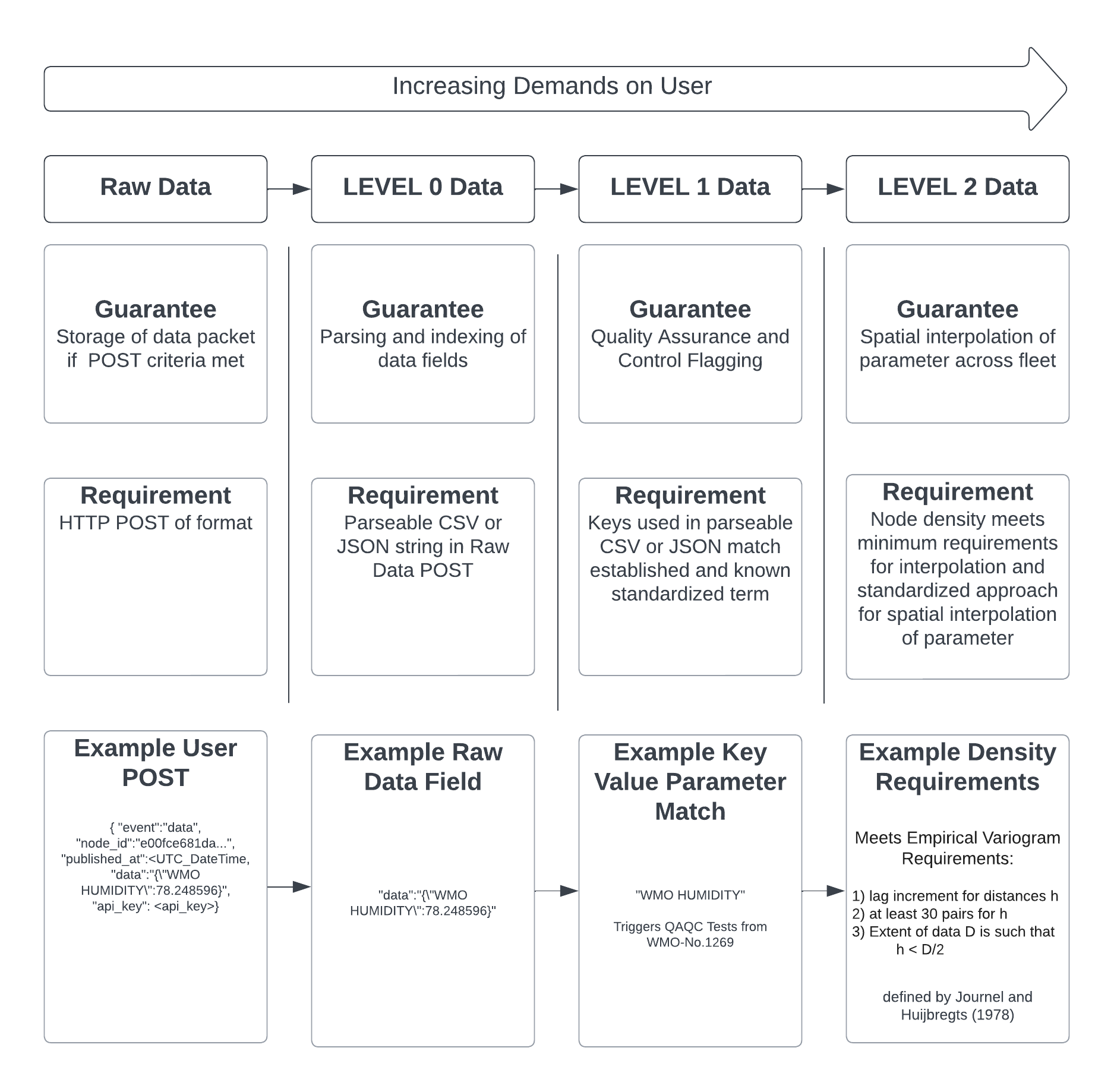}
  \caption{ Data systems are organized around a series of performance standards associated with different levels of data quality assurance and control ranging from no assurances beyond data storage through to full interpolation of a parameter across an array of nodes. As of v3, Level 1 data is implemented on database query in the Grafana dashboard and Level 2 data is done via analysts in Jupyter when needed.}
  \label{fig:fig5}
\end{figure}

\paragraph{\textit {3.3.8	Cloud Infrastructure}}\mbox{} \\
Cloud infrastructure is used throughout the system. The Particle Internet of Things platform is used for sending and receiving data from sensor nodes and for performing over-the-air firmware updates. The Google Cloud platform is used for hosting both front-end and back-end software. Through all versions of the system, Google Virtual Machines were run with Linux Debian 10+, Python 3+, Flask 1.0.3+ for API endpoint construction, and Gunicorn 20.0+ and NGINX 1.15.10+ for reverse proxy and request handling.

\section{Use Cases}
Since v0.1 of the spatial IoT systems, the stack has been used across 14 projects covering plant, animal, and natural resource sciences (Table 1). In the following section, we describe three application areas in detail.

\subsection{Irrigation Management and Monitoring}
Water quantity and quality are projected to become increasingly severe concerns for crop production and environmental quality in Minnesota \cite{hatfield_indicators_2018}. Periods of water deficit can cause crop yield loss and leave unused soil nutrients exposed to potential leaching.  Seasonal excesses of water can cause damage from in-field water logging, agricultural runoff, and water quality degradation through nutrient leaching or groundwater contamination. Climate change within the region is expected to affect the frequency and severity of these extremes. Since 2002, the use of groundwater for irrigation has increased by more than 33\% across Minnesota, such that by 2017 more than 611,000 farm acres in the state were irrigated \cite{noauthor_usda_2017}. As water management becomes increasingly important, it’s anticipated that the number of farmers using irrigation will increase even in traditionally rainfed farming areas \cite{baker_coupling_2012}. This increase in groundwater use could have a negative impact on groundwater levels and the quality of streams if not effectively managed.

Since 2016, the Irrigation Management Assistant (IMA) online tool has been adopted by over 100 regular users across Minnesota in the Little Rock Creek Groundwater area and 5-county expanded areas of Hubbard, Becker, Wadena, Otter Tail and Todd counties. These users rely on IMA to schedule irrigation for 5 different crops (corn, soybeans, alfalfa, potatoes, and edible beans) covering roughly 6,500 acres.

The technology stack described above is being used to expand the data collection network for the IMA tool and to support fundamental research on irrigation management. This includes collecting meteorological data for evapotranspiration estimation, soil moisture and temperature data for irrigation scheduling, and providing the baseline logging technology for prototype thermocouple-based temperature monitoring to parameterize bulk aerodynamic and physics-guided machine learning models. Sixteen Version 2 nodes have been repeatedly deployed on a seasonal basis since 2020, and roughly fifty additional v3 nodes are  being deployed in the summer of 2023.

\subsection{Plant Winterkill in Northern Latitudes}
Plant winterkill impacts the winter annual and perennial ground cover that society uses for recreation and ecosystem services. Golf course superintendents in northern latitudes are faced with the problem of winter damage risk every year, and undertake  cultural practices to prevent injury. Winter injury to golf course turfgrass has negative ecological impacts when perennial ground cover is absent, and economic losses when recreational activities are postponed for vegetation to re-establish in the springtime. To date, few viable solutions have been developed by the turfgrass research community. Often the specific physiological reasons for winterkill in turf systems are difficult to understand and may include the frequency and magnitude of ice encasement, gas exchange, low temperatures, desiccation, or disease. Winterkill is unpredictable and this has been due largely to the inability to capture microclimate data that characterizes these complex physiological stressors.

We performed pilot work with eight v1.0.0 Palmer systems in 2019-2020, and forty-nine v2.0.5.A and v2.0.5.B systems in 2020-2021 and 2021-2022 winters. Across node designs, node measurements included soil moisture and temperature at 3 depths; air temperature, barometric pressure, relative humidity, O2, and CO2 gas levels at the soil surface, and photosynthetically active radiation. All v1 and v2 devices were manufactured at the University of Minnesota; version 3 systems were manufactured by Caltronics Design and Manufacturing for deployment in the Fall of 2022.

\subsection{Meteorological Observations}
Multiple projects across versions have focused on meteorological observations for agriculture. This includes public deployments in Minnesota (17 nodes) and Malawi (80 nodes), as well as with private partners across the southeastern United States. We describe the public deployments in the following:

\renewcommand{\labelenumi}{\alph{enumi})}
\begin{enumerate}
\item The Minnesota Agricultural Experiment Station has been trying to develop high quality real time datasets for over 40 years. The ability to create these datasets has been limited by the cost of instrumentation, so new technologies might help alleviate this funding burden, allowing the completion of distributed weather monitoring systems. To address this challenge, we deployed 17 weather stations across Minnesota field sites consisting of both Campbell Scientific (6 nodes) and GEMS v2 equipment (11 nodes). Campbell Scientific instruments were deployed side-by-side with GEMS hardware to evaluate sensor and logger data quality. All Minnesota devices were manufactured at the University of Minnesota.
\item A comparable system was deployed across Malawi with lower cost sensors. Low-cost systems to collect data that could drive smallholder decisions in sub-Saharan Africa are of great interest to agriculture, particularly when combined with real time data sharing and market analysis. We worked with a South African-based manufacturer to produce ~80 GEMS-designed sensor nodes for this pilot project. An on-going project, we are deploying and maintaining these nodes on experimental research stations plus commercial and smallholder farms throughout Malawi, while collecting information on the robustness and perceived value of the technology within the country. All Malawi devices were manufactured by XOR in South Africa.
\end{enumerate} 

\subsection{Summary of Use Cases}
Our spatial IoT systems have enabled scientists to examine irrigation systems and management, plant winterkill, and agrometeorology. Currently, over 2,700 sensing devices have been deployed or are planned to be deployed in support of the research of over fifty scientists using these systems and to support machine learning and artificial intelligence applications in agriculture. Data generated with spatial IoT systems enabled more efficient data collection to support the generalizability of models across wider (and agriculturally variable) geographic extents.

\subsection{GEMS Sensing Service Model}
The University of Minnesota supports the creation of service organizations that advance the institution's goals of research, education, and outreach. These are heavily monitored activities to ensure mission alignment and parity with the broader external market when engaging external customers. There are two types of service organizations - internal and external - each with separate pricing. Internal sales are only allowed to cover the costs of service delivery, whereas external sales must be in line with the broader market, allowing for profit generation. Profits are to be used to further support mission-driven research, education, and outreach activities, which in the case of GEMS Sensing will be the furthering of activities in the GEMS Informatics Center related to digital agriculture.

In the IoT market for agricultural applications, there are two general business models. One involves higher upfront hardware costs and lower on-going costs for data telemetry and storage; this is common for premium logger and sensor manufacturers. The second approach involves low, or even subsidized, hardware costs and higher on-going subscription costs, and is more common among newer startup companies.

GEMS Sensing seeks to keep the barriers to entry for data collection as low as possible within the rules governing ISO and ESO sales at the University of Minnesota. For internal purchases, the hardware and on-going subscriptions associated with telemetry, data storage, and data quality assurance are provided at cost. For external customers, the hardware and on-going subscription prices are structured to incentivize public-private partnerships and generate modest revenues to support foundational research and development activities. Calendar year 2023 was the first year of offering these services for sale, and further work will consider the effectiveness of the approach. The following use cases illustrate applications of the technology across agricultural and environmental research contexts.

\section{Conclusion}
Artificial intelligence and machine learning are creating new challenges for data collection in the agricultural and environmental sciences. The need for large, clean, and analysis-ready data has created an unprecedented demand for novel data collection approaches. Open source IoT technologies are appealing because of their potential to scale data collection and enable a wide variety of use cases. However, they come with their own limitations, resulting in an agricultural IoT scaling gap. GEMS Sensing is addressing these challenges through an ecosystem of open source hardware and firmware, and standardized software and databases for storage and data quality assurance and control. Coupled with a business model focused on supporting the scaling of public science and the public-private collaborations that are increasingly common in agricultural research, the service organization supporting the use of these technologies will continue to be sustained to support real-time digital agriculture research and development.

Future developments for the GEMS Sensing systems will consider changes along the entire stack of hardware to software. For hardware and firmware, we will consider alternative modes of telemetry, including revisiting LoRa and LoRaWAN, as well as the on-going evaluation of microcontrollers for logging. Alternative logger architectures could include pivoting to a Real-time Operating System such as FreeRTOS (https://www.freertos.org/) that could reduce the costs associated with the core logger and bring firmware development in line with industry standards for embedded systems development. Further, as GEMS Sensing becomes more fully integrated as a service provider, there is a need to integrate the software stack with existing infrastructure. This includes at least partial pivots away from Google Cloud and to Minnesota Supercomputing Institute infrastructure as well as further integration with library systems such as DRUM for long term data archive (https://conservancy.umn.edu/drum), GEMS Platform for directed data sharing (https://gems.agroinformatics.org), and college-level IT for network connectivity, data storage, and data integration. 

Developing this roadmap in a way that the mission of GEMS Sensing can continue to be fulfilled while maximizing the efficiencies of working within a large institution remains an open area for development. While this sort of institution building and long-term technology planning falls outside the scope of traditional agricultural research and development, it is critical to ensure a deep and lasting impact of these technologies on agriculture for decades to come.

\section*{Acknowledgments}
We thank the many scientists and engineers who provided input and insight throughout the process including Gregg Johnson, Axel Garcia y Garcia, Matthew Bickel, Ford Denison, Forrest Izuno, Greg Cuomo, Matogen, James Meyers, Matthew Hart, Jon Hellebuyck, Ron Faber. Mention of trade names or commercial products in this publication is solely for the purpose of providing specific information and does not imply recommendation or endorsement by the U.S. Department of Agriculture or the University of Minnesota. The USDA and University of Minnesota are equal opportunity providers and employers.

\section*{Funding}
United States Golf Association
Minnesota Golf Course Superintendents Association
Michigan Turf Foundation
USDA-NIFA United States Department of Agriculture Specialty Crop Research Grant award number 2021-51181-35861
LCCMR-ENRTF 2021-266.
MAES
OVPR
V37
MnDRIVE

%Bibliography
\bibliographystyle{unsrt}  
\bibliography{IoTpaper}

\end{document}